\documentclass[prl,twocolumn,aps,preprintnumbers,floatfix,nofootinbib]{revtex4}

\usepackage{graphicx, color, dcolumn, bm, amsmath, subfigure, dsfont}

\newcommand\e{\mathrm{e}}

\newcommand{\mperp}{{m_T}}
\newcommand{\pperp}{{p_T}}
\newcommand{\rp}{{r_\perp}}

\begin{document}

\title{Collective flow in high-multiplicity proton-proton collisions
}

\author{Tigran Kalaydzhyan}
 \email{tigran.kalaydzhyan@stonybrook.edu}
\affiliation{
Department of Physics and Astronomy, Stony Brook University,
Stony Brook, New York 11794-3800, USA
}
\author{Edward~Shuryak}
\email{shuryak@tonic.physics.sunysb.edu}
\affiliation{
Department of Physics and Astronomy, Stony Brook University,
Stony Brook, New York 11794-3800, USA
}
\date{\today}

\begin{abstract}
We present an evidence of strong radial flow in high-multiplicity $pp$ collisions. We analyze the CMS data on the inclusive spectra of the charged pions, kaons and protons in the LHC $\sqrt{s}=7$\,TeV collisions. For $\langle N_{\mathrm{tracks}} \rangle \gtrsim 75$ we demonstrate the consistency of the hydrodynamic description with the (idealized) Gubser's flow. Using a one parameter fit of the model to experimental data, we obtain the initial fireball size to be of the order of 1 fm. At smaller multiplicities, the fit cannot be performed which shows a limitation of the hydrodynamic approach and provides us with falsifiability of our theory.
\end{abstract}

\maketitle

\section{Introduction}

The idea of collective effects amenable to hydrodynamic description of the proton-proton ($pp$) collisions goes all the way back to the works of Landau \cite{Landau:1953gs}. However, experimental studies of the particle spectra,
measured over decades at fixed-target and collider (ISR at CERN, Tevatron at Fermilab, RHIC at Brookhaven)  experiments, demonstrated the so-called $m_T$-scaling (to be discussed in detail below)
characteristic for individual breaking of QCD strings stretched between protons. The so-called Lund model and many forms of event generators based on this model were used to fit and explain the data.
 More recent versions of those -- such as
%
 PYTHIA 8, with a certain
 form of string interaction or
 color reconnection \cite{Sjostrand:2007gs}), successfully describe various observables associated with particle production in $pp$ collisions.
 So, $pp$ (as well as $pA$) collisions have been for long time considered qualitatively different from heavy-ion
 $AA$ collisions, for which the hydrodynamic description 
 became a mainstream since its successful explanation of RHIC data, see \cite{Heinz:2013th} and Refs. therein.

 The situation changed since the beginning of LHC operation in 2010, when CMS discovered the now famous ``ridge"
correlation  at large multiplicities (100 and higher) in $AA$ collisions (and later also in $pA$), which was confirmed to be a collective elliptic flow.
  At the same time, the Lund-model mechanism failed to describe strong growth of the mean $p_T$ with multiplicity:
  the proposed explanations were (i) appearance of radial collective flow, or (ii) increase in the parton saturation momenta $Q_s$ in GLASMA model. The former required that $T'$ slopes of the $\mperp$-spectra, defined via a fit
  \begin{align}
\frac{dN_i}{d y\,\mperp d\mperp} \sim \exp(-m_T/T') \label{eqn_T'}
\end{align}
with $\mperp=\sqrt{p_\perp^2+m^2}$, to be linearly growing with the particle mass,  $T'(m)\propto m$. The latter, however, required the slopes to be $m$-independent.

Since relevant high-multiplicity $pA$ collisions have probability of several percents, contrary to $\sim 10^{-6}$
in the $pp$ case, their studies have statistical advantage and were completed first.
 The data of the identified particle spectra ($\pi$, K, p, $\Lambda$) have clearly shown
 that the dilemma is resolved in the favor of the flow. Analysis of these data using
various versions of hydrodynamics has been made successfully, see, e.g.,
\cite{Shuryak:2013ke,Ghosh:2014eqa}.

Participation of many $N_p\sim 20$ nucleons in high-multiplicity $pA$ collisions leads to a contribution of large number of Pomeron exchanges and thus large number of produced strings. In \cite{Kalaydzhyan:2014zqa}
we discussed the string-string interaction, proposing it to be attractive one mediated by sigma-meson
exchanges, and described conditions for collectivization of the multi-string systems.

Returning to the problem of flow in the $pp$ case, let us mention that early femtoscopy data
by ALICE \cite{Aggarwal:2010aa} already included strong
evidences for such flow phenomena with a surprisingly high transverse flow velocities, as
recently revealed in the analysis by Hirono and one of us \cite{Hirono:2014dda}.
However, there were no detailed studies of the data on the identified particle spectra:
this gap we intend to fill with our paper.

%
%


\begin{figure}[t!]
\begin{center}
\includegraphics[width=70mm]{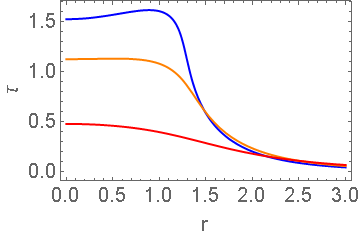}\vspace{.5cm}
\includegraphics[width=70mm]{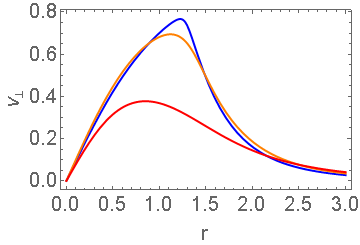}
\end{center}
\caption{ (Color online)
Profiles of the freezeout surfaces (upper plot) and the corresponding velocities (lower plot) for three different
initial system sizes: at $\epsilon_0 = 10$ and $1/q = 0.7\, \mathrm{fm}$ (upper blue), $1/q = 1\, \mathrm{fm}$ (middle orange), $1/q = 2\, \mathrm{fm}$ (lower red).}
\label{fig:profiles}
\end{figure}

\begin{figure}[t!]
\begin{center}
{\includegraphics[width=7cm]{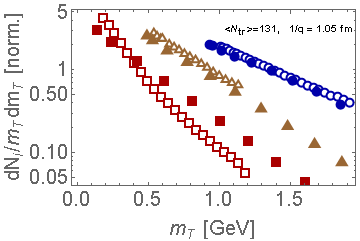}}\vspace{.5cm}
{\includegraphics[width=7cm]{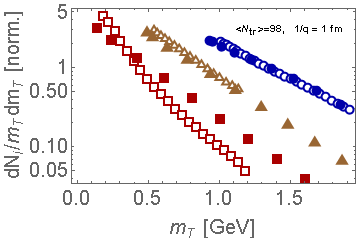}}\vspace{.5cm}
{\includegraphics[width=7cm]{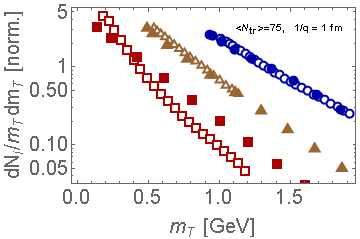}}
\caption{(Color online) Normalized spectra of pions (squares), kaons (triangles) and protons (discs) at different multiplicities.   Open symbols correspond to the CMS data \cite{Chatrchyan:2012qb} for $|\eta|<2.4$ and $\sqrt{s} = 7$ TeV, while the solid ones are obtained from the best one-parameter fit of the Gubser's flow.}
\label{fig_spectra}
\end{center}
\end{figure}

\subsection{Collective flow and spectra}

At the moment, there is basically no accepted theory of the fluctuations at the initial stage of high multiplicity $pp$ events. Therefore, like in \cite{Shuryak:2013ke}, we will use the simplest hydro solution,
 an analytic solution known as the Gubser's flow \cite{Gubser:2010ze}, which is a generalization of the Bjorken's flow for the case of finite transverse size and both radial and longitudinal expansion with respect to the beam axis.

Before we proceed, let us mention certain limitations of this approach.
(i) Since we discuss only the radial flow in this work, we deal with an axially symmetric picture. Furthermore, Gubser's flow assumes a certain initial shape of the fireball, induced by the conformal map essential for its derivation.
In reality, the shape of the actual system remains unknown, and we use this one just for practical convenience.
(ii) The  Gubser's flow assumes conformity of matter and thus EOS $\epsilon=p/3$.
Small systems, $pp, pA$, do spend most of their time in the QGP phase, and only a small fraction
of time near $T_c$ and the freezeout: for the generation of an overall flow velocity we will discuss
this approximation should be fine. (The femtoscopy radii, measured at the freezeout, are more sensitive to the final stages, and that is why we complemented Gubser's flow by a numerical solution in  \cite{Hirono:2014dda}.)
(iii) As discussed, e.g., in  \cite{Shuryak:2013ke}, the outer part of the freezeout surface
is rather unrealistic, deviating qualitatively from the more realistic hydro solutions. However, since
its contribution in the latter case is quite small, we may simply exclude this region from the consideration.

The solution is given by the energy density and transverse velocity,
\begin{align} \label{solution}
& \epsilon (\bar\tau, \bar r) = \frac{\epsilon_0 (2 q)^{8/3}}{\bar\tau^{4/3}[1+2q^2(\bar\tau^2 + \bar r^2) + q^4(\bar\tau^2 - \bar r^2)^2]^{4/3}},\\
& v_{\perp}(\bar \tau, \bar r) \equiv \tanh \kappa (\bar \tau, \bar r) = \frac{2 q^2 \bar \tau \bar r}{1 +q^2 \bar\tau^2 +q^2 \bar r^2},
\end{align}
where $\kappa (\bar \tau, \bar r)$ is radial flow rapidity; $\bar r$ and $\bar \tau=\sqrt{t^2-z^2}$ are the radial coordinate and the (longitudinal) proper time, respectively. The solution is parameterized by a pair $\left(q, \epsilon_0\right)$.
The dimensionless energy density parameter $\epsilon_0$ is related to the entropy per unit rapidity,
\begin{align}
 \epsilon_0 = f_\ast^{-1/3}\left(\frac{3}{16 \pi} \frac{dS}{d \eta}\right)^{4/3},
\end{align}
where $f_\ast = 11$ is the number of effective degrees of freedom in quark-gluon plasma \cite{Gubser:2010ze}.
The entropy per unit rapidity is given by the charged particle multiplicity,
\begin{align}
 \frac{dS}{d \eta} \simeq 7.5 \frac{d N_{\rm ch}}{d \eta}.
\end{align}

  Parameter $q$ characterizes an inverse transverse size of the system at the beginning of the hydrodynamic phase.
Since, as we already emphasized, there is no theory of the initial state, we do not know its value \textit{a priori}.
Our study of the spectra can thus be seen as an attempt to find its value from the data,
using the radial flow phenomenon.

To simplify our further calculations, we consider the Gubser's solution in dimensionless variables $\tau=q \bar{\tau}, \,\,\,r=q \bar{r}$,
\begin{align}
 & {\epsilon \over q^4}  =  \frac{{\epsilon}_0\, 2^{8/3}}{\tau^{4/3}\left[1+2(\tau^2 +
r^2)+(\tau^2-r^2)^2\right]^{4/3}}\,,\label{epsilongubser}\\
 & v_\perp(\tau, r) \equiv \tanh \kappa (\tau, r) = \frac{2 \tau r }{1 + \tau^2 + r^2}\,.
\end{align}
To illustrate typical flow solutions, we depict them in Fig.~\ref{fig:profiles} for some of the parameters similar to the ones used in this paper. The freezeout surface profile is obtained by solving (\ref{epsilongubser}) for $\epsilon = f_*\,T_{\mathrm{f}}^4$ and $T_\mathrm{f}=170$ MeV. As one can see from the upper plot, for the lower curve the freezeout happens from the edge inwards, meaning the system cools down gradually. However, the upper curve indicates a strong radial flow, i.e. the system undergoes a fast expansion due to a high internal pressure and then suddenly freezes out. In other words, a plateau in the freezeout profile (and a ``knee'') is an indicator of a strong radial flow.
The absolute magnitude of the radial flow, shown in the lower part of Fig.~\ref{fig:profiles}, one can see the
radial distribution of the corresponding flow velocities.

In order to obtain inclusive particle spectra, we use the Cooper-Frye formula \cite{Cooper:1974mv},
\begin{align}
\frac{dN_i}{d y\, p_{T} d p_{T} d \phi_p}=  \int \frac{p^\mu d^3 \sigma_\mu(x)}{(2 \pi)^2} f_i(x, p),
\label{eq:CP}
\end{align}
where $\phi_p$ is the azimuthal angle of $\vec p_{T}$,
$f_i(x, p)$ is the distribution function for the particles of the chosen type $i$, and the integration is performed over a hypersurface of constant temperature (the freezeout temperature, $T_\mathrm{f}=170$ MeV).
For the further discussion one should also introduce the so-called ``transverse mass'', $m_T\equiv \sqrt{p_T^2 + m^2}$, with a useful property $m_{T}\, d m_{T} = p_{T}\, d p_{T}$.

In the Boltzmann approximation, $f_i=g_i\,\e^{(p\cdot u - \mu_i)/T}$ (for kaons and protons), and an azimuthally symmetric case, the Eq.~(\ref{eq:CP}) reduces to \cite{Heinz:2004qz}
\begin{align}
&   \frac{dN_i}{d y\,\mperp d\mperp} =
   \frac{g_i}{\pi^2} \int_0^{r_{\mathrm{cut}}} \rp d\rp\,\tau\, \e^{\mu_i(\rp)/T_f}   \times \nonumber\\
&   \times\biggl[
   \mperp {\rm K}_1\Bigl(\frac{\mperp\cosh\kappa(\rp)}{T_f}\Bigr)
          {\rm I}_0\Bigl(\frac{\pperp\sinh\kappa(\rp)}{T_f}\Bigr)
\nonumber\\
&   -\pperp\frac{\partial\tau}{\partial\rp}
   {\rm K}_0\Bigl(\frac{\mperp\cosh\kappa(\rp)}{T_f}\Bigr)
   {\rm I}_1\Bigl(\frac{\pperp\sinh\kappa(\rp)}{T_f}\Bigr)\biggr],\label{longformula}
\end{align}
where $g_i$ is a number of states for the given particle mass $m_i$, and $\tau$ is taken at the freezeout surface.

We already mentioned that the outer tail of the Gubser's solution is clearly unphysical: its power fall off
with  distance is different from the exponentially falling nuclear densities. We simply do not include the
part outside a peak value (it either corresponds to small times, when hydro regime is not yet developed, or describes hydro incorrectly), and take a cut-off $r_{\mathrm{cut}}$, which is defined by the position of integrand's maximum as a function of $r_\perp$. In what follows, we assume the chemical potential $\mu_i=\mathrm{const}$ and normalize the distributions, which makes the exponent and other numerical prefactors irrelevant.
One should not expect to reproduce pion spectra in this approximation because of the Bose-Einstein statistics and resonance decays which are not taken into account. For pions one should, in principle, change the Eq.~(\ref{longformula}) by multiplying arguments of the exponent and Bessel functions by $n$, multiply the whole expression by $(-1)^{n+1}$ and sum over $n\in \mathds{N}$, but in our case it turned out to be not essential and does not change the result much, so we present it as it is. This procedure would take into account the Bose-Einstein statistics but not the other effects.

\begin{table}[b]
\centering
\begin{tabular}{| c || c | c | c | c | c | c |}
\hline
$\langle N_{\mathrm{tracks}} \rangle$ &  $\epsilon_0$ & \,$1/q$ [fm]\, & $dS/d\eta$& \,$v_\perp^{\mathrm{max}}$\, & $T'(p)$ & $T'(K)$ \\ \hline \hline
131  	&  12.7  &  1.05$\pm$0.05    & 204.7     &  0.71 & 574 MeV & 397 MeV  \\\hline
98 	    &  8.6   &  1.00$\pm$0.05    &  153.1    &  0.68 & 458 MeV & 338 MeV \\\hline
75 	    &  6.0   &  1.00$\pm$0.05    &  117.2    &  0.63 & 394 MeV & 301 MeV \\
\hline
\end{tabular}
\caption{Parameters used in the calculation and the output.}\label{paramtable}
\end{table}

\begin{figure}[t!]
\begin{center}
\includegraphics[width=70mm]{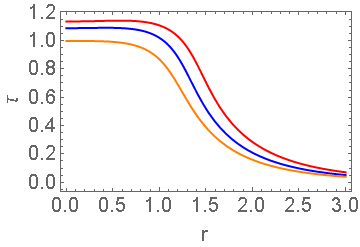}\vspace{.5cm}
\includegraphics[width=70mm]{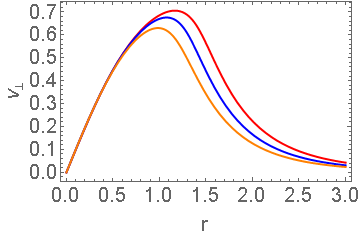}
\end{center}
\caption{ (Color online)
Profiles of the freezeout surfaces (upper plot) and velocities (lower plot) for the Gubser solutions at various values of the parameters $(1/q\, [\mathrm{fm}], \epsilon_0)$ used in our calculations. Upper red - (1.05, 12.7), middle blue - (1.0, 8.6), lower orange - (1.0, 6.0).}
\label{fig:ourprofiles}
\end{figure}

Before turning to the results, let us discuss some qualitative features  of the spectra.  First, let us assume for a moment
a complete  absence of the flow, i.e. put $\kappa=0$. Then the Eq.~(\ref{longformula}) reduces to (\ref{eqn_T'})
meaning the $m_T$ spectra for all hadrons are identical, with a slope $T'=T_\mathrm{f}$ (the so-called, $m_T$-scaling).
By turning on the flow, $\kappa\neq 0$, one would violate this $m_T$-scaling, since distributions for different particles will in general have different shapes, and in particular different slopes.  Such difference is indeed visible from the data in Figure~\ref{fig_spectra}. The larger is the multiplicity, the more pronounced is the $m_T$-scaling violation. It is clear that $\langle N_\mathrm{tracks}\rangle=75$ is a transition case, i.e. slopes are nearly similar. It is also visible from the corresponding Gubser solution, Fig.~\ref{fig:ourprofiles}, where this case is almost similar to a complete absence of the radial flow.

Results of our calculations  are shown in Fig.~\ref{fig_spectra}, together with the experimental data.
We use the CMS data \cite{Chatrchyan:2012qb} for charge particles distributions in $\sqrt{s}=7$ TeV $pp$ collisions. The transverse momentum (or mass) distributions for $\sqrt{s}=0.9$ and $2.76$ TeV in the chosen multiplicity classes, if present, are practically the same and we do not consider them separately. Typical parameters and output are listed in Table~\ref{paramtable}. It is important to note that for lower multiplicities, i.e. $\langle N_{\mathrm{tracks}}\rangle < 75$, we could not perform any fit, which would describe the data. We treat this fact as a breakdown of hydrodynamic approach for low multiplicities. It is amusing to note that collective effects start appearing at a similar multiplicity in $pPb$ collisions \cite{Chatrchyan:2013nka, Kalaydzhyan:2014zqa}.

 For the nonrelativistic region, $m_T \sim m_i$, the inverse slope parameter $T'$ characterizes a blueshifted freezeout temperature \cite{Scheibl:1998tk}, i.e. $T' = T_\mathrm{f} + m_i\langle v_\perp \rangle^2/2$. As one can see from Table~\ref{paramtable}, the inverse slope for protons and kaons extracted from experimental data is, indeed, larger than the freezeout temperature.
The fact that the pion spectra are steeper (especially at the low transverse momenta) can be related to the
so-called feed-down pions, the presence of additional pions from resonance decays. Such contribution to other species is much smaller and, therefore, we considered a simultaneous fit to the kaon and proton slopes
to be the priority. As seen from the plot, this goal is reached.

In Fig.~\ref{fig:ourprofiles} we show the freezeout surface profile and the corresponding transverse velocity distribution, 
producing the aforementioned spectra. Note that the value of the scale parameter happen to be very close
to $q =1$ fm in all cases. Looking at the upper plot Fig.~\ref{fig:ourprofiles} one finds that both the
proper time and radius of the fireball {\em at the freezeout} are close to 1 fm. Looking at the lower plot of  Fig.~\ref{fig:ourprofiles} one finds that the maximal value of the transverse velocity at its edge reaches
$v_\perp^{max}\approx 0.65$ or so.

\section{Conclusions and discussion}

We demonstrated that the high multiplicity $pp$ collisions, like those in $pA$, possess strong indications for a
collective radial flow. The magnitude of the flow, needed to explain the spectra of identified secondaries, mostly kaons and protons, is quantified. We further observe that the freezeout time and radius of the system
are both close to 1 fm. As we already mentioned in the introduction, independent analysis of the femtoscopy data
 \cite{Hirono:2014dda} provides similar velocity estimates, and even a bit smaller size $1/q\approx 2/3 \,\mathrm{fm}$.
To use a hydrodynamic description one should make sure
the mean free path in QGP 
is much smaller than that size. This conclusion, following from the data, is of course highly nontrivial.
One may also wonder how for such a small  system it was possible to acquire
 the  transverse velocity as large as $v_\perp^{max}\approx 0.65$
(at its edge). Hydrodynamics, in the particular form of Gubser flow solution, provides a picture of that
as a space and time dependence of the energy density.
As time $\tau$ goes to zero, one sees that the corresponding energy density (\ref{solution}) becomes very large,
and it is physically obvious that at some ``initial time" $\tau_i$ the hydrodynamical description
should break down.  From our analysis we, of course, do not know what this value can be, since the final observable does not  depend on it.

 We also do not know, and do not even speculate, what physical process is responsible for the system formation.
 Let us only comment on the string model interpretation put forward for the $pA$ data in our paper
    \cite{Kalaydzhyan:2014zqa}. In our analysis, we considered the initial system as a collection of strings,
stretched between the colliding proton and a nucleus, and then undergoing a collective collapse.
    We also introduced there the so-called diluteness of the ``spaghetti" of the QCD strings (i.e. fraction of the transverse area occupied by strings), which was about 0.3 or so, i.e.
    small enough to treat the system as sparse.

      However, in the $pp$ case we considered in this paper, there is no large number
 $\sim 20$ of participant nucleons and Pomerons, and we have no clue whether
 a muti-string description can or cannot be used at all. All we can say is that both systems, high multiplicity $pp$ and $pA$
 collisions, have very similar femtoscopy sizes and flow magnitudes.

\acknowledgements This work is supported in part by the U.S. Department of Energy under Contract No. DE-FG-88ER40388.

\end{document}